\documentclass[10pt,a4paper, onecolumn]{article}
\usepackage[latin1]{inputenc}
\usepackage[T1]{fontenc}
\usepackage[english]{babel}
\usepackage[]{csquotes}
\usepackage{cite}
\usepackage{hyperref}
\usepackage{epigraph}
\usepackage[]{graphicx}
\usepackage{microtype} 	
\usepackage{vmargin}		
\usepackage{color}
\usepackage{amssymb}
\usepackage{amsmath}
\usepackage{latexsym}
\usepackage{amsthm}
\usepackage{bbm}
\usepackage{array}
\usepackage{braket}
\usepackage{microtype}
\usepackage{color}
\usepackage{float}

\newcommand{\Tphoto}{{\Delta T}}
\newcommand{\Tmean}{\Tphoto_{\text{mean}}}
\newcommand{\Tvar}{\Tphoto_{\text{var}}}

\setmargnohfrb{20mm}{20mm}{20mm}{20mm}
\title{\bf \Large Photon number statistics uncover the fluctuations in non-equilibrium lattice dynamics}

\author{\normalsize Martina Esposito$^{1, \dagger}$,  Kelvin Titimbo$^{1,2, \dagger}$, Klaus Zimmermann$^{1, 2}$, Francesca Giusti$^1$, Francesco Randi$^1$,\\
\normalsize Davide Boschetto$^3$, Fulvio Parmigiani$^{1,4}$, Roberto Floreanini$^2$, Fabio Benatti$^{1,2,*}$ and Daniele Fausti$^{1,4,*}$\\
\\
\small ${}^1$Dipartimento di Fisica, Universit\`a di Trieste,
34127 Trieste, Italy\\
\small ${}^2$Istituto Nazionale di Fisica Nucleare, Sezione di Trieste, 34014 Trieste, Italy\\
\small ${}^3$ Ecole polytechnique, ENSTA ParisTech, CNRS, Universit\'e Paris-Saclay,\\
\small 828 bd des Mar\'echaux, 91762 Palaiseau cedex, France\\
\small ${}^4$Sincrotrone Trieste S.C.p.A., 34127 Basovizza, Italy\\
\small ${}^\dagger$ These authors contributed equally to the work\\
\small ${}^*$ Corresponding authors.
}

\begin{document}
\pagestyle{plain}
\footskip = 30pt
\maketitle
\textbf{
Fluctuations of the atomic positions are at the core of a large class of unusual material properties ranging from quantum para-electricity to high temperature superconductivity. Their measurement in solids is the subject of an intense scientific debate focused on seeking a methodology capable of establishing a direct link between the variance of the atomic displacements and experimentally measurable observables. Here we address this issue by means of non-equilibrium optical experiments performed in shot-noise limited regime. The variance of the time dependent atomic positions and momenta is directly mapped into the quantum fluctuations of the photon number of the scattered probing light. A fully quantum description of the non-linear interaction between photonic and phononic fields is benchmarked by unveiling the squeezing of thermal phonons in $\alpha$-quartz.}\\


In a classical description the displacement of the atoms along the vibrational eigenmodes of a crystal can be measured with unlimited precision. Conversely, in the quantum formalism positions and momenta of the atoms can be determined simultaneously only within the boundary given by the Heisenberg uncertainty principle. For this reason, in real materials, in addition to the thermal disorder, the atomic displacements are subject to fluctuations which are intrinsic to their quantum nature. While various evidences suggest that such quantum fluctuations may be of relevance in determining the onset of intriguing material properties, such as quantum para-electricity, charge density waves, or even high temperature superconductivity \cite{sachev00, nozawa01, newns07, hashimoto12, grilli96, muller91, cohen15}, the possibility of measuring quantum fluctuations in solids is the subject of an intense debate \cite{HuNori96, HuNori96a, garret97, garret97a, HuNori99,  Misochko00, Bartels00, Hussain10, sauer10, reiter11, misochko11, hu11, Misochko13, Riek15}.

The time evolution of atomic positions in materials is usually addressed by means of non-equilibrium optical spectroscopy. An ultrashort light pulse (the pump) impulsively perturbs the lattice and a second one (the probe), properly delayed in time, measures a response that is proportional to the spatially-averaged instantaneous atomic positions. In those experiments, the time dependent atomic displacements are often revealed by an oscillating response, commonly dubbed \textit{coherent phonon} response\cite{Dhar94,jhonson09,Henighan15,Trigo13,merlin97,papalazarou12,li13,weiner91, zeiger92, lobad01, nori97}, at frequencies characteristic of the vibrational modes of the material.
In this framework, it has been shown that a non-linear light-matter interaction can prepare non classical vibrational states \cite{Misochko00, HuNori96} such as squeezed states, where the fluctuations of the lattice position (or momentum) can fall below the thermal limit. A reduction below the vacuum limit is known as vacuum squeezing \cite{sauer10}.

Here we propose a joint experimental and theoretical approach to access the fluctuations of the atomic positions in time domain studies. An experimental apparatus that allows for the measurement of the photon number quantum fluctuations of the scattered probe pulses in a pump and probe setup is adopted. The connection between the measured photon number uncertainty and the fluctuations of the atomic positions is given by a fully quantum mechanical theoretical description of the time domain process. Overall we prove that, under appropriate experimental conditions, the fluctuations of the lattice displacements can be directly linked to the photon number quantum fluctuations of the scattered probe pulses. Our methodology, that combines non-linear spectroscopic techniques with a quantum description of the electromagnetic fields, is benchmarked on the measurement of phonon squeezing in $\alpha$-quartz.

\section*{Results}
\subsection*{Shot-noise limited pump and probe experiments}
In the non-linear spectroscopy formalism, the excitation mechanism of phonon states in transparent materials is called impulsive stimulated Raman scattering (ISRS) \cite{weiner91}. The susceptibility tensor $\chi^{(3)}$ connects the induced third order polarization $P^{(3)}$ to three fields: $\mathcal{E}_{1}(\omega_1)$, $\mathcal{E}_{2}(\omega_2)$ and  $\mathcal{E}_{3}(\omega_3)$ \cite{muk}. In conventional two pulses pump and probe experiments, the fields  $\mathcal{E}_2(\omega_2)$ and $\mathcal{E}_3(\omega_3)$  are two different frequency components of the pump laser pulse. In particular, all photon pairs  such that $\omega_3 - \omega_2 = \Omega$, where $\Omega$ is the frequency of the Raman active vibrational mode, contribute to ISRS \cite{merlin02}. The interaction of the probe field $\mathcal{E}_1(\omega_1)$ with the photo-excited material induces an emitted field, $\mathcal{E}_{\text{EF}} (\omega)$ which depends on the pump-probe delay and carries information about the specific Raman mode excited in the crystal. Experimental details are reported in (Supplementary Note 1).
We choose a polarization layout designed to excite $E$-symmetry Raman optical modes in $\alpha$-quartz at room temperature and get an emitted field with polarization orthogonal to the probe one \cite{run2006} (Supplementary Fig. 1).

\begin{figure}[p]
\centering
\includegraphics[]{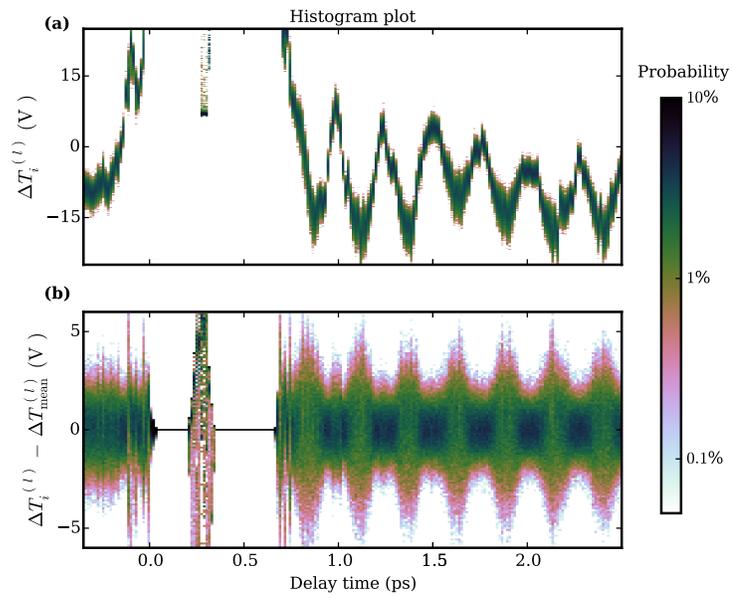}
\caption{\textbf{Time domain transmittance histogram plot.} $\Tphoto_i$ distribution as a function of pump-probe delay for a representative scan $l$. (a) For each time delay a color plot of the normalized histogram of $N=4000$ acquired pulses is shown. (b) Histogram plot of $\Tphoto_i$ centered at zero. The data shown are obtained with the largest pump fluence used in the experiments ($25 \,  \text{mJ} \, \text{cm}^{-2}$).}
\label{fig1}
\end{figure}

We propose here a new approach to time domain studies.
The experimental layout is similar to standard pump and probe experiments.
The sample is excited by an ultrashort pump pulse and the time evolution of the response is measured by means of a second much weaker probe pulse, that interacts with the photo-excited material at a delay time $\tau$. The unique characteristics of our setup are: i) unlike standard experiments, where the response is integrated over many repeated measurements, our system can measure individual pulses; ii) the apparatus operates in low noise conditions allowing for the measurement of intrinsic photon number quantum fluctuations. In detail, we adopt a differential acquisition scheme where each probe pulse is referenced with a second pulse which has not interacted with the sample.  For each measurement the differential voltage is digitized and integrated, giving the transmittance $\Tphoto_{i}$ for the $i\textsuperscript{th}$ measurement.
For every given pump and probe delay $\tau$, we repeat this single pulse measurements for $N=4000$ consecutive pulses. Fig.~\ref{fig1} (a) gives a useful visual representation of the obtained data. For one pump and probe scan $l$ the normalized histogram of $N=4000$ acquired pulses for each delay time is shown.
Each histogram represents the distribution of the measured $\Tphoto_i$ for a specific delay time $\tau$.
For a clearer visualization of the physically meaningful information in the time evolution of the statistical distribution, Fig.~\ref{fig1} (b) reports the histogram centered at zero.

The pump and probe scan is repeated several times and each $l\textsuperscript{th}$ scan provides $\Tmean^{(l)}=\frac{1}{N}\sum_i{\Tphoto_i}$, and  $\Tvar^{(l)}=\frac{1}{N}\sum_i{[\Tphoto_i-\Tmean^{(l)}]^2}$. Finally the averages of these two quantities are calculated over all $M$ scans as $\Tmean = \frac{1}{M}\sum_{l=1}^{M}\Tmean^{(l)}$ and $\Tvar = \frac{1}{M}\sum_{l=1}^{M}\Tvar^{(l)}$.

\begin{figure}[p]
\centering
\includegraphics[]{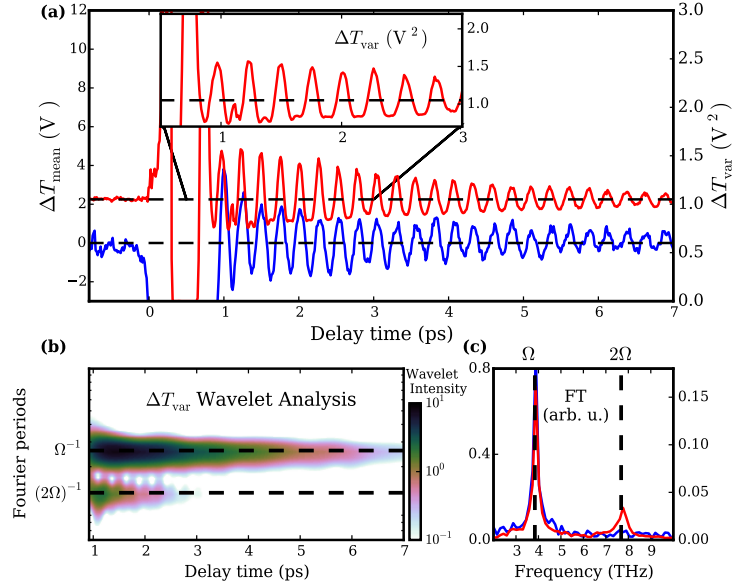}
\caption{\textbf{Time domain transmittance mean and variance.} (a) $\Delta T_{\text{mean}}$ (blue curve) and $\Delta T_{\text{var}}$ (red curve) as a function of the pump-probe time delay. The zero time is the instant in which pump and probe arrive simultaneously on the sample. In the inset a zoom of the variance for the first $3$ ps is shown. (b) Wavelet analysis (Morlet power spectrum) of the variance oscillating part. (c) Fourier transforms of the oscillating parts of mean (blue curve) and variance (red curve). In (a) and (c) the left axis is related to the mean while the right axis is related to the variance.}
\label{fig2}
\end{figure}

The time domain response, averaged over $M = 10$ scans, is shown in Fig.~\ref{fig2} (a) for a representative pump fluence of $14 \,  \text{mJ} \, \text{cm}^{-2}$ (a pump fluence dependent study is reported later).  The blue curve depicts the time evolution of the mean value of the transmittance $\Tmean$, whereas the red curve shows the time evolution of its variance $\Tvar$. The Fourier transform of the mean (Fig.~\ref{fig2} (c), blue curve) has a single peak which is ascribed to the E-symmetry quartz vibrational mode with frequency $\Omega = 128 \, \text{cm}^{-1} = 3.84 \, \text{THz} $ \cite{porto67}. The same frequency component is observed in the Fourier transform of the variance (Fig.~\ref{fig2} (c), red curve). In addition, a second peak at twice the phonon frequency appears exclusively in the variance. A wavelet analysis of the variance oscillations allows for a time domain study of the two frequency components (Fig.~\ref{fig2} (b)):  one notices that, while the fundamental frequency survives for roughly $7 \, \text{ps}$, the $2 \Omega$ component vanishes within the first $2 \, \text{ps}$. The different lifetimes between the $\Omega$ and $2 \Omega$ components of the variance are seen also by a close inspection of the raw data distribution plotted in Fig.~\ref{fig2} (b).

Note that the $2 \Omega$ in our data is visible only in experimental conditions where the noise is dominated by the quantum uncertainty, a situation which is known as shot-noise regime. In such conditions $\Tvar$ measures the quantum variance of the scattered probe photon number. A full characterization of the detection system is reported in (Supplementary Note 2), including the shot-noise characterization (Supplementary Fig. 2) and the analysis of classical noise sources (Supplementary Fig. 3 and Supplementary Fig. 4). It should further be stressed that in experimental conditions where the noise is larger and dominated by classical sources the $2 \Omega$ contribution to the noise becomes unmeasurable.

The presence of the $2 \Omega$ frequency component is suggestive of phonon squeezing, as it has been indicated by Raman tensor models \cite{HuNori96, HuNori96a,HuNori99}. Nevertheless, the experimental evidences up to date lack a direct comparison with a reliable quantum noise reference \cite{garret97, garret97a, Misochko00, Misochko13}. Hence, in these experiments the observation of the $2 \Omega$ frequency in the optical noise is considered as an indication of phonon squeezing, but not an unequivocal proof.
In details, a $2 \Omega$ oscillating optical noise was reported in \cite{Misochko00}, but later ascribed to an artifact \cite{Hussain10} due to the experimental amplification process. In particular, it has been demonstrated that amplification artifacts become more relevant when, using a lock-in amplifier based acquisition, the time constant of the lock-in increases with respect to the time between steps in the pump-probe delay. This gives rise to maxima in the noise where the derivative of the mean signal is maximal \cite{Hussain10}. Here we use a pump power density which is almost $3$ orders of magnitude higher than in \cite{Misochko00, Hussain10}. In addition, we observe a $2 \Omega$ frequency component in the optical variance which exhibits maxima in correspondence with the minima of the derivative of the mean signal, hence ruling out possible artifacts of the kind described in \cite{Hussain10}.

\subsection*{Fully quantum description of impulsive stimulated Raman scattering}

In order to predict how the fluctuations of the atomic positions in a lattice can be mapped onto the photon number quantum fluctuations of the probe field, we develop a novel theoretical approach to time domain studies which treats quantum mechanically both the material and the optical fields involved in the non-linear processes. Several semiclassical models describe the possibility of generating "classical" (coherent states) and non classical vibrational states by photo-excitation.
In particular, for transparent materials like quartz, the most commonly used approach is to adopt Raman tensor models where the interaction between photons and phonons is not mediated by dipole allowed electronic transitions. In this condition, interactions linear in the phonon operators allow for the generation of coherent vibrational states, while high order interactions are required for the generation of non classical squeezed states \cite{HuNori99, nori97,HuNori96a}.
In materials with allowed dipole transitions, as in presence of excitons, different models based on electron-phonon coupling Hamiltonians have been proposed. In those models it has been shown that squeezed phonon states can result only by successive excitations with a pair of pulses \cite{sauer10, reiter11}. All these models mainly adopt semiclassical approaches where the optical fields are described classically \cite{muk}, and therefore are unable to reproduce the quantum proprieties of the probe optical field that can be measured with the shot-noise limited pump and probe setup presented here. The key aspect of our approach, allowing us to bridge this gap, is to study both generation and detection of phonon states using a fully quantum formalism through an effective photon-phonon interaction, which is descriptive of experiments in transparent systems, such as $\alpha$-quartz. The basic tool is a quantum Hamiltonian able to describe both pump and probe processes. Being linear and bilinear in the photon and phonon operators, this Hamiltonian accounts for the possible generation of coherent and squeezed phonon states through the pump process. In particular, it models also the detection of the photo-excited phonon states, describing the probing process by a fully quantum approach, providing in this way a direct comparison with the experimentally measured photon number quantum fluctuations of the scattered probe pulses \cite{titimbo15}.

In this framework the first step is to adopt a quantized description for the mode-locked pulsed laser fields \cite{esposito14}. Each mode of frequency $\omega_j = \omega_0 + j \delta$, where $\omega_0$ is the pulse central frequency, $\delta$ is a constant depending on the laser repetition rate and $j$ is an integer, is quantized and described by single mode creation and annihilation operators $\hat{a}_j^{\dagger}$ and ${\hat{a}_j}$. In this framework ISRS can be modelled by means of an effective impulsive interaction Hamiltonian which is descriptive of both the pumping and the probing processes. In both processes two optical fields with orthogonal polarizations (denoted with subscript $x$ or $y$)  are involved: two pump fields in the pumping process and the probe and the emitted field in the probing process. The interaction Hamiltonian has the form
\begin{eqnarray}\label{Hint}
\mathcal{H}  =  \sum_{j, j'=-J}^{J} \left[ g^{1}_{j, j'} \,  \mu_{\text{d}}  \big(\hat{a}_{xj}^{\dagger} \, \hat{a}_{yj'} \, \hat{b}^{\dagger} + \hat{a}_{xj} \, \hat{a}_{yj'}^{\dagger} \, \hat{b} \big) +  \, g^{2}_{j, j'} \,  \mu_{\text{s}}   \big( \hat{a}_{xj}^{\dagger} \, \hat{a}_{yj'} \, (\hat{b}^{\dagger})^2 + \hat{a}_{xj} \, \hat{a}_{yj'}^{\dagger} \, \hat{b}^{2} \big) \right],
\end{eqnarray}
where $2J +1$ is the total number of modes within a mode-locked optical pulse, $\hat{b}$ and $\hat{b}^{\dagger}$ are the phonon annihilation and creation operators, $\mu_{\text{d}}$ and $\mu_{\text{s}}$ are coupling constants and the functions $g^{\ell}_{j, j'}$ take into account the relations between the frequencies of the involved fields,
\begin{equation}
\nonumber
g^{\ell}_{j, j'} =
\left\{
\begin{array}{rl}
1 & \mbox{if } j' = j + \frac{\ell \Omega}{\delta} \\
0 & \mbox{elsewhere},
\end{array}
\right.\quad \ell=1, 2\ ,
\\
\end{equation}
with $\Omega$ the phonon frequency.  A complete interaction Hamiltonian should contain also second order processes involving phonons with opposite momenta. However, since the probe detects only the $\mathbf{k}\simeq0$ optical transition, we can make use of an effective Hamiltonian that accounts only for this kind of process.

\begin{figure}[p]
\centering
\includegraphics[]{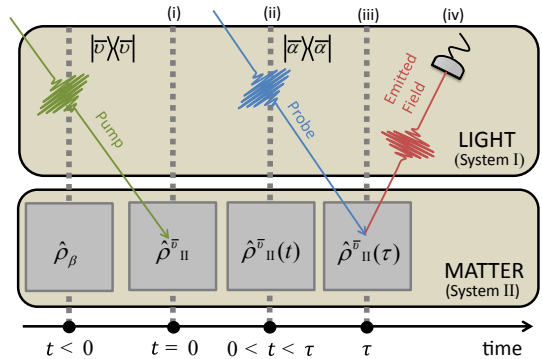}
\caption{\textbf{Sketch of the four steps effective theoretical model}. The steps are indicated with roman numbers. The details of the theory for each step are reported in the text. The photon and phonon system are denoted with I and II, respectively.}
\label{fig3}
\end{figure}

The whole theoretical description of the experiment can be rationalized in a four step process as sketched in Fig.~\ref{fig3}: (i) generation of phonon states in the pumping process, (ii) evolution of the produced vibrational state, (iii) probing process and (iv) read out of the emitted photon observables.

(i) Initially, the sample is in thermal equilibrium and it is described by a thermal phonon state $\hat{\rho}_{\beta}$, at inverse temperature $\beta$. The laser pump pulse is described by a multimode coherent state of high intensity $\ket{\bar{\nu}} = \ket{\nu_{-J}} \otimes \cdots \otimes \ket{\nu_{J}}$, where $\ket{\nu_j}$ are single mode coherent states associated with all the frequency components within the  pulse. Each $\ket{\nu_j}$ is an eigenstate of the annihilation operator $\hat{a}_j$  of photons in the mode of frequency $\omega_j$, $\hat{a}_j\ket{\nu_j}=\nu_j\ket{\nu_j}$. We indicate with $\bar{\nu}$ the vector whose components are the amplitudes $\nu_j$. The equilibrium (pre-pump)  photon-phonon  state $\hat{\rho} = \ket{\bar{\nu}}\bra{\bar{\nu}} \, \otimes \, \hat{\rho}_{\beta}$ is  instantaneously  transformed  into $\hat{\rho}^{\bar{\nu}} = \mathcal{U} \, \hat{\rho} \, \mathcal{U}^{\dagger}$ by  means  of the  unitary  operator $\mathcal{U} = \text{exp}\{-i\mathcal{H} \}$. Since the pumping operator $\mathcal{U}$ acts on a high intensity photon coherent state $\bar{\nu}$, we can use the mean field approximation for the photon degrees of freedom and replace $\hat{a}$ with $\nu$ and $\hat{a}^{\dagger}$ with $\nu^{*}$ for both pump modes involved in equation \eqref{Hint}, thus replacing $\mathcal{U}$ by
\begin{eqnarray}
&&\hskip-.5cm
\mathcal{U}_{\bar{\nu}} =\exp\Bigl\{ -i \sum_{j, j'=-J}^{J} \big[ g^1_{j, j'} \,  \mu_{\text{d}}  \big({\nu}_{xj}^{*} \, {\nu}_{yj'} \, \hat{b}^{\dagger} + {\nu}_{xj} \, {\nu}_{yj'}^{*} \, \hat{b} \big) \, + \,g^2_{j, j'} \,  \mu_{\text{s}}   \big( {\nu}_{xj}^{*} \, {\nu}_{yj'} \, (\hat{b}^{\dagger})^2 + {\nu}_{xj} \, {\nu}_{yj'}^{*} \, \hat{b}^{2} \big)\big]\Bigr\}.
\label{Unu}
\end{eqnarray}
The evolution operator generates  coherent and squeezed phonon states, respectively, through the linear and quadratic terms in the phonon operators $\hat{b}$ and $\hat{b}^{\dagger}$. The initial state $\hat{\rho}^{\bar{\nu}}$ contains information about both photons and phonons. Tracing over the photon  degrees  of  freedom, the resulting  state  $\hat{\rho}^{\bar{\nu}}_{\text{II}}$ describes the excited phonons  brought out of equilibrium by the impulsive pump process.

(ii) The time evolution of the excited phonons is described by using an open quantum systems approach,  namely  by  means  of  a  suitable  master equation of  Lindblad form \cite{alicki, breuer} that  takes  into  account,  besides the quantum unitary evolution, also the dissipative and noisy effects due to the interaction with a thermal environment.

(iii) The incoming probe pulses are in the multimode coherent state $\ket{\bar{\alpha}}$. The  probing  process at time $\tau$ is  described  by  the same impulsive unitary operator $\mathcal{U}$ used for the pump.  However, in this case we can apply the mean field approximation only to the probe photon operators with $x$ polarization, which correspond to a much more intense field than those with $y$ polarization. Moreover, since the probe field is much weaker than the pump field, the quadratic terms in the interaction Hamiltonian in equation \eqref{Hint} can now be neglected. The resulting unitary operator is
\begin{equation}
\label{Ualpha}
\mathcal{U}_{\bar{\alpha}'} = \exp\{-i \lVert \bar\alpha'\rVert \big(\hat A(\bar\alpha') \, \hat{b}^{\dagger} + \hat A^\dag(\bar\alpha') \, \hat{b} \big)\}
, \quad \hat A(\bar\alpha')= \frac{1}{\lVert \bar\alpha'\rVert}\sum_{j=-J}^{J} (\alpha'_j)^{*}  \, \hat{a}_{yj}\  ,
\end{equation}
where $\bar\alpha'$ is the vector with components $\alpha'_j = \, \mu_{\text{d}} \,\sum_{j'=-J}^{J}\,  g^1_{j', j} \,{\alpha}_{xj'}$ and $\hat A(\bar\alpha')$ is a collective photon annihilation operator such that $\left[\hat A(\bar\alpha'), \hat A^\dag(\bar\alpha') \right] = 1$.\\
 The latter unitary operator acts on  a  state  of  the  form $\ket{\bar{\alpha}}\bra{\bar{\alpha}} \, \otimes \, \hat{\rho}^{\bar{\nu}}_{\text{II}} (\tau)$. The information  about  the phonons are extracted by  measuring the emitted field photons. In particular, the emitted photon state $\hat{\rho}_{\text{I}} (\tau)$ is obtained by tracing away the phonon degrees of freedom.

\begin{figure}[p]
\centering
\includegraphics[]{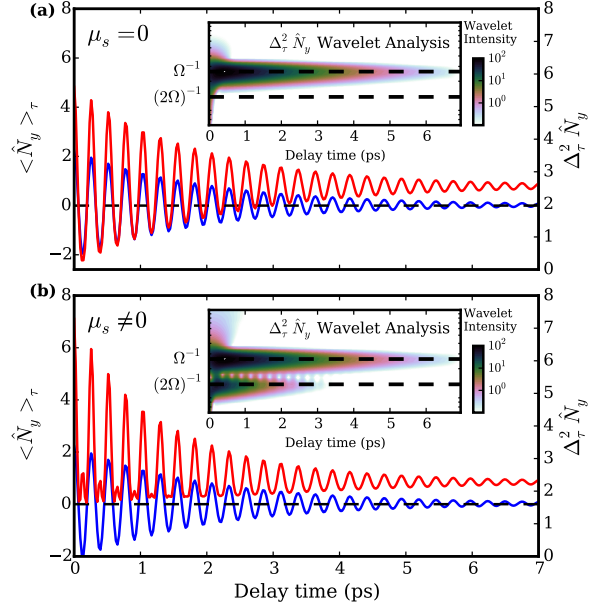}
\caption{\textbf{Model predictions}. Theoretical calculations of the mean value and the variance of the number of photons of the emitted field. The left axis is related to the mean while the right axis is related to the variance. Comparison between the case with squeezing coupling constant $\mu_{\text{s}}=0$ (a) and $\mu_{\text{s}} \ne 0$ (b). A wavelet analysis (Morlet power spectrum) of the variance is reported in the inset for both cases.}
\label{fig4}
\end{figure}

(iv) The possible quantum features of the phonon state, e.g. squeezing, can be read off as they are imprinted into $\hat{\rho}_{\text{I}} (\tau)$. In particular for each time delay $\tau$ we can compute the quantities $\braket{\hat N_y}_\tau = \braket{\hat A^\dag(\bar\alpha') \hat A(\bar\alpha')}_\tau$ and $\Delta^2_\tau\hat N_y =\braket{\hat{N}_y^2}_\tau - \braket{\hat{N}_y}_\tau^2 $, which correspond to the observables measured in the experiment, that are the mean value and the variance of the number of photons of the emitted field.
The details of the theoretical model are reported in (Supplementary Note 3). The theoretical results for $\mu_{\text{s}}=0$ and $\mu_{\text{s}} \ne 0$ are shown in Fig.~\ref{fig4} together with the corresponding wavelet analysis for the variance of the number of emitted photons. The calculations reproduce the experimental results in Fig.~\ref{fig2}, revealing a $2 \Omega$ frequency component in the variance, only when the pump creates squeezed phonon states ($\mu_{\text{s}} \ne 0$). In particular, for  $\mu_{\text{s}} \ne 0$, the model reproduces the different lifetimes between the $\Omega$ and $2 \Omega$ components in the variance observed in the experiments. The explicit expressions for the theoretically predicted amplitudes of both the frequency components in the variance are reported in (Supplementary Note 3), showing that the same damping constant, characterizing the dissipative phonon time evolution, contributes differently to the two frequency components giving rise to different decay times.
\section*{Discussion}
\begin{figure}[p]
\centering
\includegraphics[]{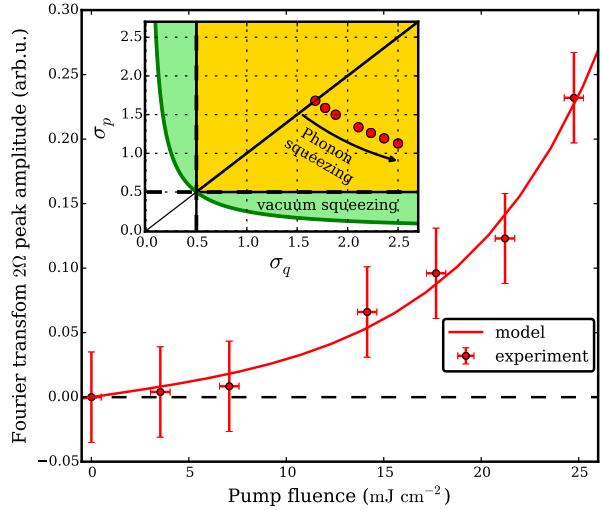}
\caption{\textbf{Fourier transform $2 \Omega$ peak amplitude of the variance.} Amplitude of the $2  \Omega$ peak of the Fourier transform of the time dependent variance $\Tvar$. The error bars indicate the standard deviation over $10$ scans. Comparison between experiments and theory as a function of the pump fluence. In the inset the uncertainties for the phonon position and momentum operators, calculated from the model, are plotted for the corresponding pump fluences. }
\label{fig5}
\end{figure}

The proposed effective interaction model is further validated by a pump fluence dependence study. Fig.~\ref{fig5} shows the amplitude of the $2 \Omega$ peak in the Fourier transform of the variance, $\Tvar$, as a function of the pump fluence.
A fluence dependence study of the $\Omega$ peak is reported in (Supplementary Fig. 5). The functional behaviour obtained from the model predictions (continuous line in Fig.~\ref{fig5}) agrees with the experimental data only in presence of a pump-induced squeezing of the phonon mode ($\mu_{\text{s}} \neq 0$). The increase of the $2 \Omega$ peak amplitude with the pump fluence allows us to give a direct estimation of the uncertainties of the phonon conjugated quadratures which are reported in the inset of Fig. \ref{fig5} for the different excitation fluences (calculation details are given in (Supplementary Note 3)). For high pump fluences the uncertainty on one of the phonon quadratures falls below the thermal limit at the equilibrium, indicating the squeezed nature of the photo-excited thermal vibrational states.
Our novel experimental approach allows for the direct measurement of the photon number quantum fluctuations of the probing light in the shot-noise regime and our fully quantum model for time domain experiments maps the phonon quantum fluctuations into such photon number quantum fluctuations, thereby providing an absolute reference for the vibrational quantum noise. The comparison of the predicted noise with the experimental photon number quantum uncertainty, measured in shot-noise conditions, allows us to unveil non classical vibrational states produced by photo-excitation.
A future extension of the model taking into account the role of the electronic degrees of freedom would allow to extend such a study from transparent materials to complex absorbing systems.

In conclusion, a Raman active phonon mode has been impulsively excited via ISRS in a $\alpha$-quartz by means of a pump and probe transmittance experiment with single pulse differential acquisition in noise conditions limited by intrinsic probe photon number fluctuations. A fully quantum mechanical effective model where both phonons generation and detection are studied through the same effective coupling Hamiltonian establishes a direct connection between the measured photon number quantum fluctuations of the emitted probe field and the fluctuations of the atomic positions in a real material. Our novel approach is used here to reveal distinctive quantum properties of vibrational states in matter, in particular the squeezed nature of photo-excited phonon states in $\alpha$-quartz.
Finally, we stress that our innovative approach paves the way for future studies addressing the role of unconventional vibrational states in complex systems\cite{newns07, muller91}, and the thermodynamics of vibrational states \cite{campisi11, goold15} possibly in the quantum regime.

\section*{Acknowledgments}
The authors are grateful to John Goold, Roberto Merlin, Keith Nelson, Mauro Paternostro and Charles Shank for the insightful discussions and critical reading of the manuscript. We thank the \textit{CAEN} company for the project and the realization of the differential detector used in the experiments. We acknowledge Giovanni Franchi for the design of the detector and Riccardo Tommasini for support during its development. The experimental activities have been carried out at the TRex labs within the Fermi project at Trieste's synchrotron facility. This work has been supported by a grant from the University of Trieste (FRA 2013) and a grant from Italian Ministry of Education Universities and Research MIUR (SIR 2015, Controlling quantum Coherent Phases of matter by THz light pulses).
\section*{Author contributions}
M.E., F.G., F.R., D.B. and D.F. performed the experiments. F.R. and M.E. developed the acquisition system. K.T., K.Z., R.F. and F.B. developed the theoretical description. M.E., F.G. and D.F. analyzed the experimental data, M.E, D.F., F.B. and F.P. coordinated the project and wrote the manuscript with contributions from all the co-authors. D.B. proposed the $\alpha$-quartz case of study and provided the sample. The experiment has been conceived by D.F and F.P.

\section*{Competing financial interests}
The authors declare no competing financial interests.

\DeclareGraphicsRule{.ai}{pdf}{.ai}{}

\end{document}


\pagestyle{plain}
\footskip = 30pt

\renewcommand{\figurename}{\textbf{Supplementary Figure}}

\leavevmode\vfill
\begin{figure}[H]
\centering
  \includegraphics[width=0.6\textwidth]{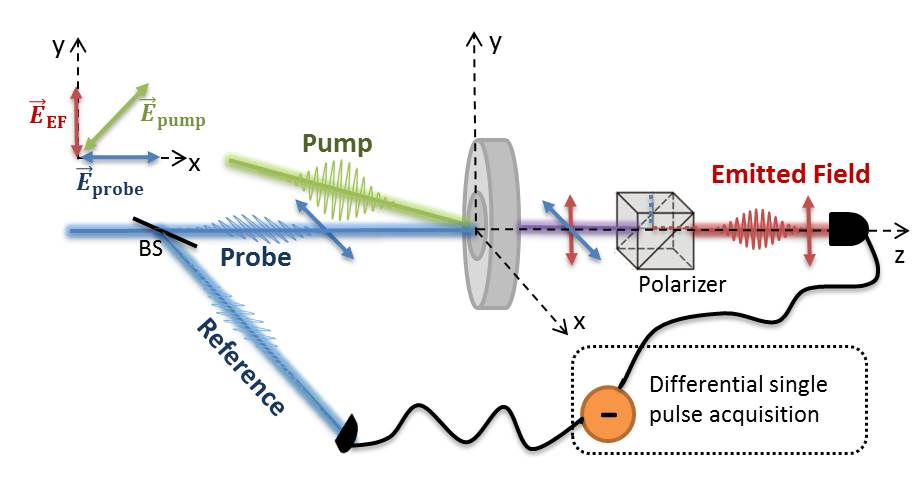}
  \caption{\textbf{Scheme of the experimental setup.} The sample is depicted at the origin of the coordinates. The polarization configuration of pump, probe and emitted field is indicated on top-left. After the interaction with the sample the probe undergoes a polarization selection in order to detect the emitted field only. The differential acquisition system is also sketched.}
  \label{fig1SM}
\end{figure}
\vfill
\newpage
\leavevmode\vfill
\begin{figure}[H]
\centering
  \includegraphics[width=0.55\textwidth]{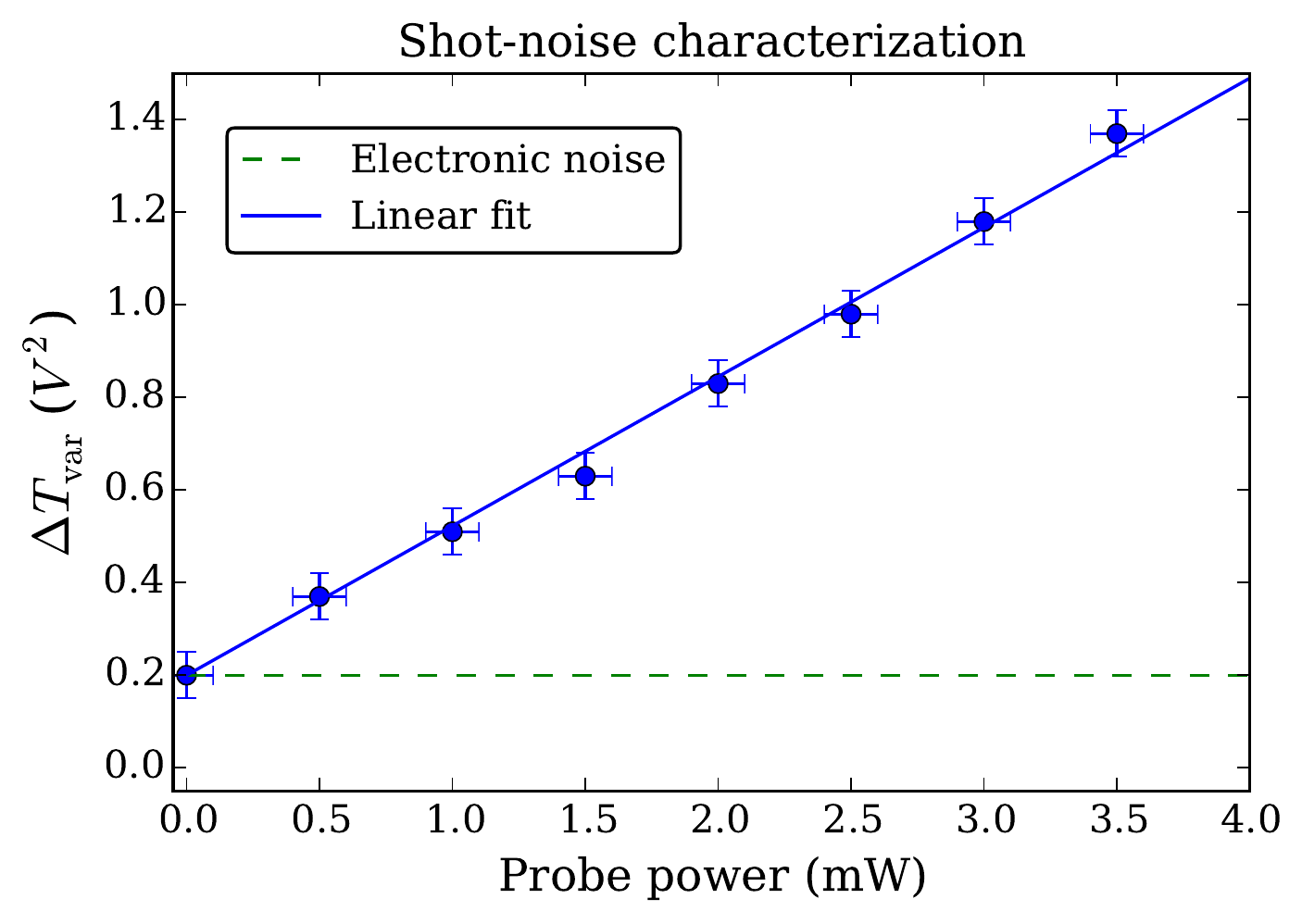}
  \caption{\textbf{Shot-noise characterization of the detection apparatus}. Variance of $4000$ acquired differential pulses as a function of the probe power. The vertical error bars indicate the standard deviation over $10$ repeated measurements, the horizontal error bars indicate the instrumental error of the power-meter.}
  \label{fig2SM}
\end{figure}
\vfill

\newpage

\leavevmode\vfill
\begin{figure}[H]
\centering
  \includegraphics[width=0.55\textwidth]{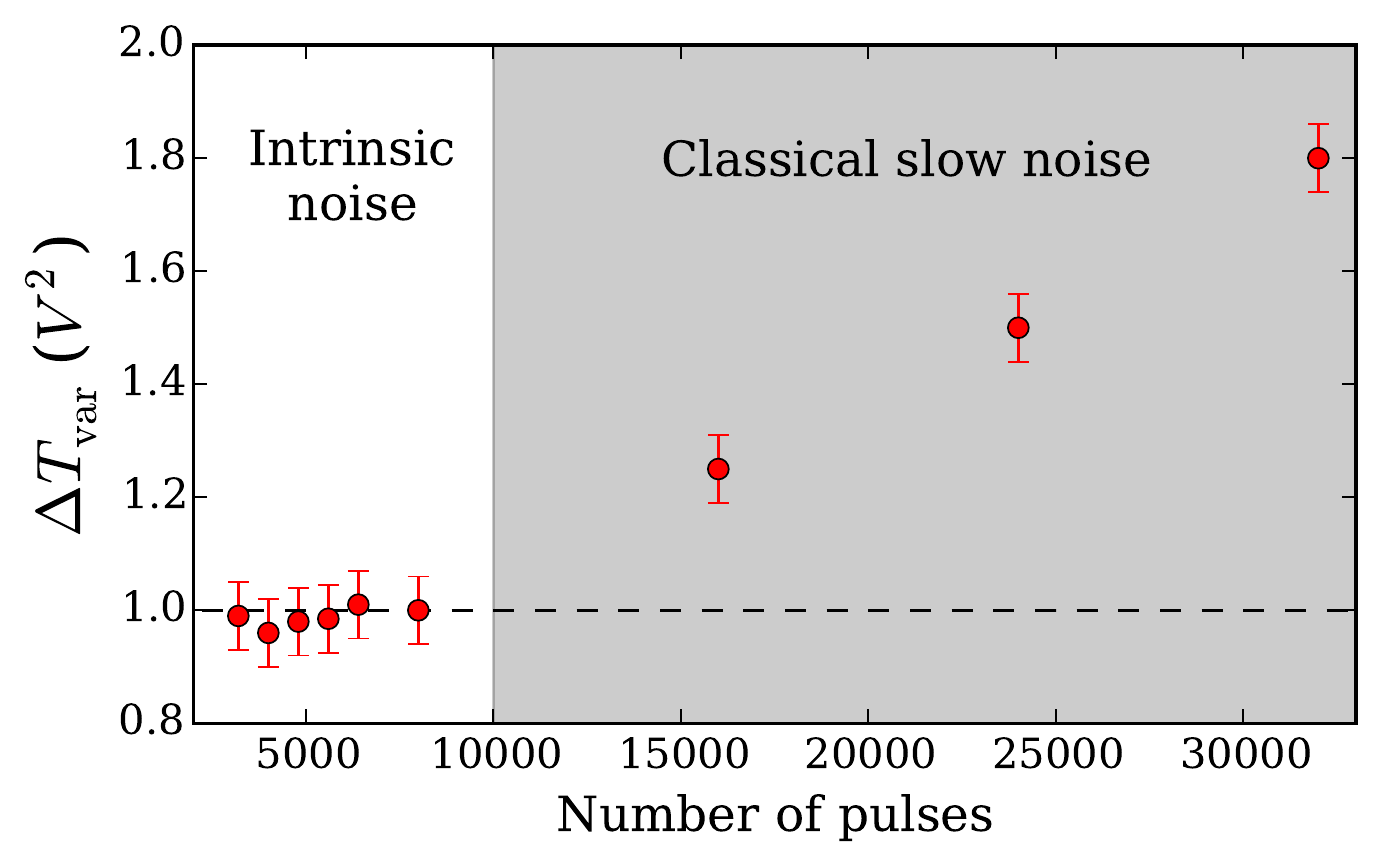}
  \caption{\textbf{Noise characterization as a function of the number of acquired pulses.} Variance, $\Delta T_{\text{var}}$, in absence of the pump, as a function of the number of acquired successive pulses. The dashed line indicates the shot-noise level. The grey area represents the region in which classical slow noise sources contributes to the variance. The error bars indicate the standard deviation over $10$ repeated measurements.}
  \label{figNpulses}
\end{figure}
\vfill

\newpage

\leavevmode\vfill
\begin{figure}[H]
\centering
  \includegraphics[width=0.55\textwidth]{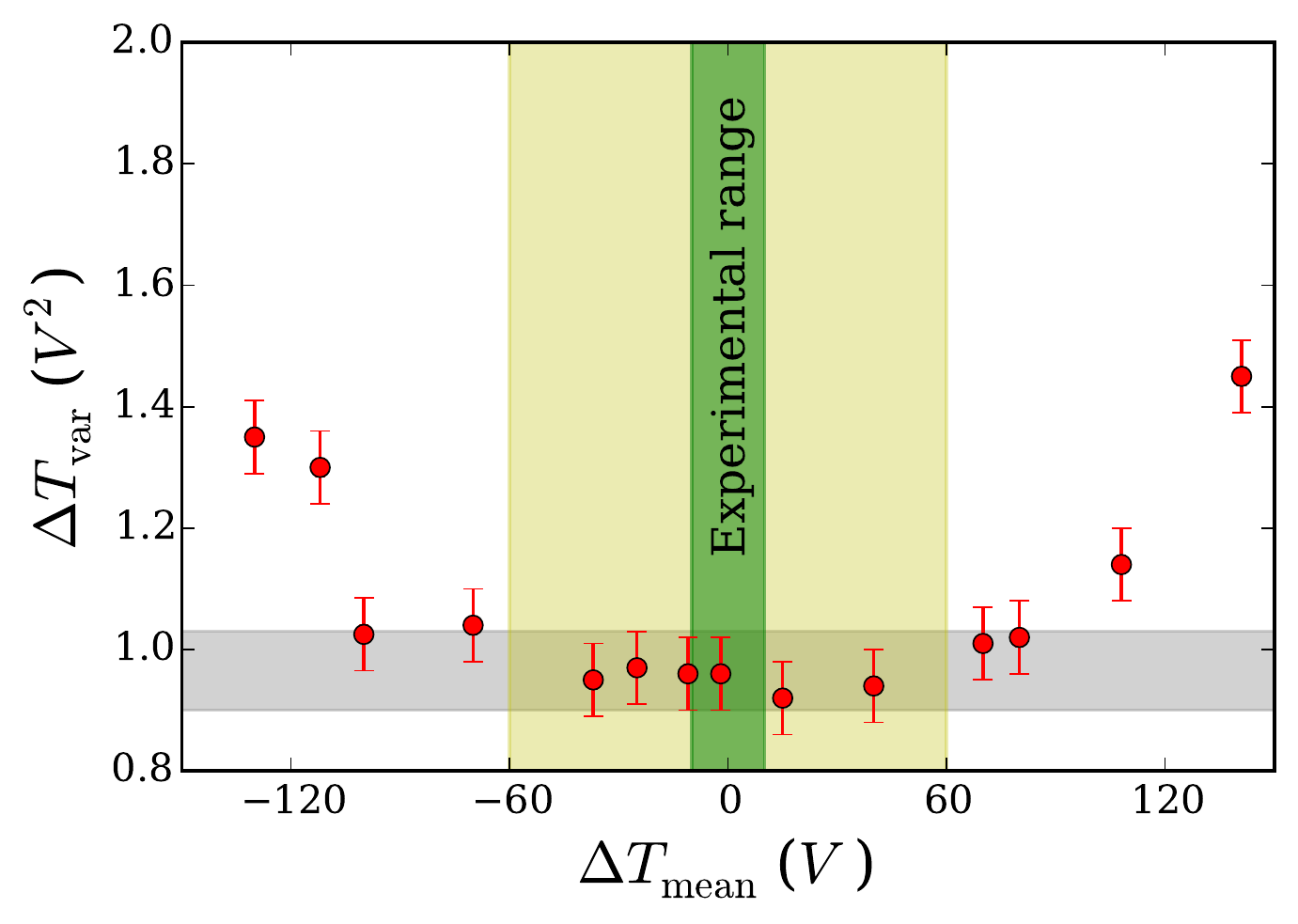}
  \caption{\textbf{Noise characterization as a function of the unbalance.} Variance, $\Delta T_{\text{var}}$, in absence of the pump for a fixed number of pulses ($N=800$), as a function of the unbalance between the transmitted probe pulse and the reference pulse, expressed as acquired mean voltage $\Delta T_{\text{mean}}$. Grey area: range of minimal noise; yellow area:  unbalance region in which the noise randomly fluctuates in a range of minimal values;  green area: unbalance region in which the experiments have been performed. Notice that we use the convention of expressing the voltage acquired for every single pulse acquisition as the sum of the voltages digitized for $500$ points at $1~\text{GS}\, \text{s}^{-1}$. For reference, the digitized measurement of a differential pulse with a $1~\text{V}$ ($-1~\text{V}$) voltage peak corresponds to a value of $300~\text{V}$ ($-300~\text{V}$). The error bars indicate the standard deviation over $10$ repeated measurements.}
  \label{figUnbalance}
\end{figure}
\vfill

\newpage

\leavevmode\vfill
\begin{figure}[H]
\centering
  \includegraphics[width=0.55\textwidth]{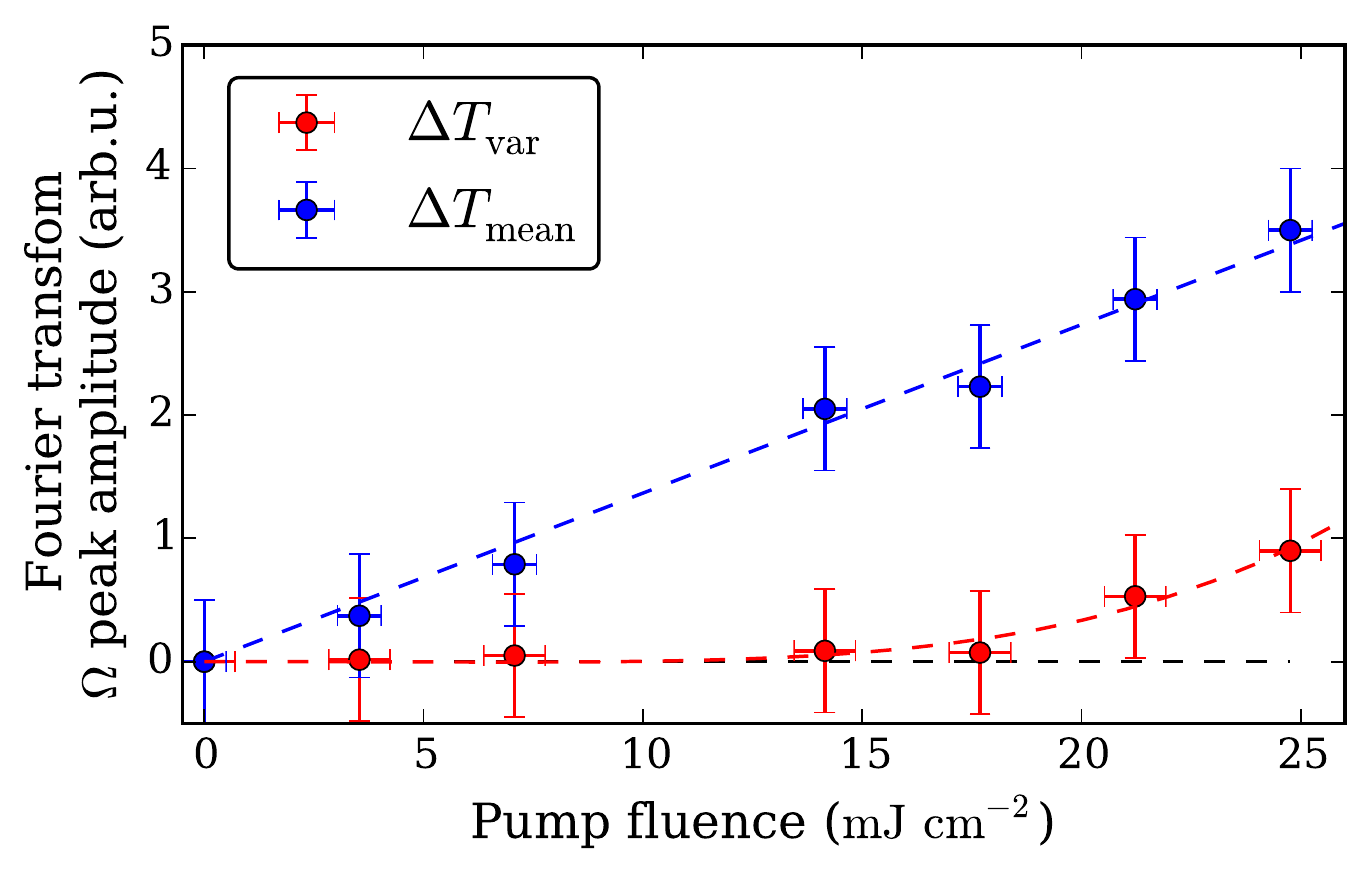} 
  \caption{\textbf{ Amplitude of the $\Omega$ peak of the Fourier transform}. Fourier transform $\Omega$ peak of the time dependent mean, $\Delta T_{\text{mean}}$, (blue points) and of the time dependent variance, $\Delta T_{\text{var}}$, (red points). The dashed lines are a guide for the eyes. The error bars indicate the standard deviation over $10$ scans.}
  \label{figOmega}
\end{figure}
\vfill
\newpage

\section*{\large Supplementary Note 1. Polarization dependent ISRS in quartz}

Impulsive stimulated Raman scattering (ISRS) is a non resonant excitation mechanism of lattice vibrations in transparent materials by ultrashort laser pulses. It is a four-wave mixing process due to third order polarization effects. The Cartesian components of the third order non-linear polarization are given by
\begin{eqnarray}
P_{i}^{(3)} (\omega) =\sum_{jkl} \, \chi_{ijkl}^{(3)} \, [\mathcal{E}_{1}( \omega_1)]_{j} \, [\mathcal{E}_{2}( \omega_2)]_{k} \, [\mathcal{E}_{3}( \omega_3)]_{l},	
\end{eqnarray}
where $\chi_{ijkl}^{(3)}$ is the susceptibility tensor, $\mathcal{E}_{1}$ is the probe field and  $\mathcal{E}_{2}$ and  $\mathcal{E}_{3}$ are the pump fields. The susceptibility tensor determines the polarization selection of vibrational modes that can be excited via ISRS \cite{etche87}.\\
In particular, quartz Raman active vibrational modes are $4$ totally symmetric modes of symmetry $A_1$ and $8$ doubly degenerate modes of symmetry $E$ (transverse and longitudinal) \cite{porto67}.

In our experiment, the sample is a $1 \, \text{mm}$ thick $\alpha$-quartz, oriented in order to have the principal symmetry axis parallel to the probe propagation direction. A scheme of the chosen experimental geometry  is shown in Supplementary Figure~\ref{fig1SM}.
Both pump and probe come from the same laser source, a $250 \, \text{kHz}$ mode-locked amplified Ti:Sapphire system. The pulse duration is $80 \,  \text{fs}$, the fractional change in the probe transmission due to the pump is of the order of $5\%$ for  a pump fluence of $25 \,  \text{mJ} \, \text{cm}^{-2}$. The excited phonon state is detected via the scattering of the probe pulse which arrives on the sample with time delay $\tau$ with respect to the pump. The transmitted light undergoes a polarization selection through a polarizer positioned after the sample.

The pump direction is almost collinear with the probe one. Assuming that the involved optical fields propagate along the $z$ direction, we can limit our analysis to the $xy$ plane. In this case the quartz Raman tensors assume the form \cite{umari01}
\begin{equation}
	\begin{array}{ccc}
	A=\begin{pmatrix} a & 0 \\ 0 & a \end{pmatrix} & E^{T}=\begin{pmatrix} c & 0 \\ 0 & -c \end{pmatrix} & E^{L}=\begin{pmatrix} 0 & -c \\ -c & 0 \end{pmatrix}.
	\end{array}
\label{RT}
\end{equation}
Following the notation in \cite{run2006}, the susceptibility tensor can be expressed as:
\begin{equation}
\chi^{(3)}_{ijkl} = A_{ij} \, A_{kl} + E^{T}_{ij} \, E^{T}_{kl} + E^{L}_{ij} \, E^{L}_{kl},	
\end{equation}
where each index can assume the values $1, 2$ associated to the direction $x$ and $y$ respectively. Thus, the susceptibility tensor $\chi^{(3)}_{ijkl}$ gives rise to a $4\times4$ block-matrix
\begin{equation}
\begin{pmatrix}
\begin{pmatrix} a^2 + c^2 & 0 \\ 0 & a^2 - c^2 \end{pmatrix} & \begin{pmatrix} 0 & c^2 \\ c^2 & 0 \end{pmatrix}\\
 & & \\
\begin{pmatrix} 0 & c^2 \\ c^2 & 0 \end{pmatrix} & \begin{pmatrix} a^2 - c^2 & 0 \\ 0 & a^2 + c^2 \end{pmatrix}	
\end{pmatrix}.
\label{chi}
\end{equation}
The first two indexes refer to the outer matrix elements and describe the polarization components of the emitted field ($i$ index) and of the probe field ($j$ index), while the last two indexes ($k$ and $l$) indicate the inner  matrix elements and describe the polarization components of the two pump fields.

In particular, we are interested  in selecting the excitation of an $E$ symmetry Raman mode. For this purpose we use a probe linearly polarized along $x$ and we perform a polarization selection after the sample in order to detect the emitted field component orthogonal to the probe (along $y$). This polarization configuration allows the selection of the susceptibility matrix elements  $\chi^{(3)}_{21kl}$ associated with the involved $E$ phononic mode. Notice that such elements vanish when $k=l$ that is when the two pump fields are both polarized along $x$ or along $y$. Thus, in order to activate the process, we need the two pump fields (two frequency components of the same laser pulse) to have orthogonal polarizations. This is possible when the pump pulse is linearly polarized along a direction in between $x$ and $y$. In particular, the efficiency of the ISRS is maximal when the pump polarization is at $45\,^{\circ}$ with respect to the $x$ axis. This is indeed the configuration we chose and consequently the matrix elements involved  in our experiment are $\chi^{(3)}_{21kl (k \neq l)} = c^2$, getting an emitted field almost collinear with the unscattered probe photons and with polarization orthogonal to the probe one. We configure a polarizer after the sample in order to transmit the emitted field polarization only. The global polarization configuration is sketched in Supplementary Figure~\ref{fig1SM}.

Note that a full extinction of the unscattered probe is experimentally not achievable (polarizer extinction rate $10^5$). The residual probe acts as a local oscillator amplifying the emitted field within the total signal \cite{muk}.


\newpage
\section*{\large Supplementary Note 2. Single pulse differential acquisition system in shot-noise limited regime}

The acquisition system is made of a balanced amplified differential photodetector and a fast digitizer (\textit{Spectrum} M3i.2132-exp) with sampling rate $1~\text{GS}\, \text{s}^{-1}$. The differential photodetector consists of two \textit{Hamamatsu} \textit{S3883} Silicon PIN photodiodes with $0.94$ quantum efficiency connected in reverse bias  and followed by a  low-noise charge amplifier. The photo-currents generated by the two photodiodes in response to a single optical pulse impinging on them (transmitted probe pulse on the first photodiode and reference pulse on the second photodiode) are physically subtracted and the resulting charge is amplified using \textit{CAEN} custom designed electronic components. In particular, the preamplifier sensitivity is $5.2~\text{mV}\,\text{fC}^{-1}$ with a linear response up to about $2~\text{V}$ of pulse peak voltage. Note that we use here the convention of expressing the voltage acquired for every single differential pulse acquisition, $\Delta T_i$ , as the sum of the voltages digitized for $500$ points at $1~\text{GS}\, \text{s}^{-1}$. For reference, the digitized measurement of a pulse with a $1~\text{V}$ voltage peak corresponds to a value of $300~\text{V}$.

Notice that our acquisition apparatus distinguishes itself by avoiding the lock-in amplification and the possible artifacts associated with its use \cite{Hussain10}: the single pulse differential acquisition system we adopted allows the individual measurement of each single transmitted probe pulse, and gives access to the intrinsic photon-number quantum fluctuations.

In order to distinguish the intrinsic noise (shot-noise) from other contributions, we tested the experimental set-up in absence of the pump. The variance $\Delta T_{\text{var}}$ of $ 4000$ differential pulses is measured for different powers of the probe. The optical noise, shown in Supplementary Figure \ref{fig2SM},  is linear with a constant offset representing the electronic noise. This behavior is characteristic of the shot-noise regime \cite{bach}.  We chose a probe power within the linearity interval ($2.5~\text{mW}$, $0.2~\text{mJ}\, \text{cm}^{-2}$ on the sample; note that this corresponds roughly to $10^6$ photons per pulse scattered in cross polarization on the detector). This provides a reference for the shot noise value of about $1~\text{V}^2$ which is used to benchmark the noise in time domain experiments and avoid additional noise sources.\\
The detector shot-noise linearity test demonstrates that we are sensible to quantum fluctuations of the photon number. In particular, the shot-to-electronic noise at the maximum probe intensity in the linear regime is approximately $10$~dB.

We further characterized our experimental apparatus by measuring the variance, $\Delta T_{\text{var}}$, in absence of the pump,  as a function of the number of acquired pulses. The results are shown in Supplementary Figure \ref{figNpulses}. We observe that the variance increases when the number of  acquired successive pulses increases, this means that for long acquisitions a slow noise contribution makes higher the measured variance (grey area in Supplementary Figure \ref{figNpulses}). Thus, we chose to acquire $N = 4000$ successive pulses per step in order to guarantee the statistical meaningfulness of the data but at the same time avoiding contributions of classical slow noises. Note that the time to time noise in the mean number of photons reported in the main text is larger than the value of the variance.
This is explained by considering that the fluctuations in the mean values are made large by slow noise of classical nature (the differences between the measurements scan by scan). On the contrary the variance is dominated by intrinsic fluctuations and, as discussed in the main text, it is calculated for every scan separately and then averaged. The results reported in  Supplementary Figure \ref{figNpulses} show that acquiring up to about $10^4$ pulses guarantees that classical slow noise contributions are excluded.

Moreover we measured the obtained variance $\Delta T_{\text{var}}$ for a fixed number of pulses ($N=800$) in absence of the pump as a function of the unbalance between the transmitted probe pulse and the reference pulse, that is as a function of the acquired mean voltage $\Delta T_{\text{mean}}$. The results of such characterization measurement are shown in Supplementary Figure \ref{figUnbalance}. One can notice that the noise randomly fluctuates in a range of minimal values (grey area) for small positive or negative unbalance (yellow area). For larger unbalance the noise starts to increase due to artifacts in the amplification process. All the time domain experiments reported here have been performed within the region of small detector unbalancing (green area) in order to be sure of working in shot noise limited conditions.

The results of our time resolved experiments show the presence of a $2 \Omega$ frequency component only in the variance of the probe photon number. A pump fluence dependent study of the of the amplitude of the $2 \Omega$ peak in the Fourier transform of the variance is reported in the main text. For completeness we report here the amplitude of the $\Omega$ peak in the Fourier Transform of the mean, $\Delta T_{\text{mean}}$, and of the variance, $\Delta T_{\text{var}}$, as a function of the pump fluence. The data are show in Supplementary Figure \ref{figOmega}.

\newpage
\section*{ \large Supplementary Note 3. An effective fully quantum mechanical model for ISRS}

The effective fully quantum mechanical approach to ISRS followed in the main text consists in the pump process, the subsequent dissipative, irreversible phonon dynamics and the probe process, all of them being described by quantum dynamical maps \cite{titimbo15}.

Before being hit by the pump laser beam described by photons in a multi-mode coherent
state  $\ket{\bar{\nu}}\bra{\bar{\nu}}$, the state of the relevant phonon mode at frequency $\Omega$ is appropriately taken to be a thermal state at inverse temperature $\beta$
\begin{equation}
\label{thermal state}
\hat{\rho}_\beta=\left(1-{\rm e}^{-\beta\Omega}\right)\,{\rm e}^{-\beta\Omega\,\hat b^\dag \hat b}\ .
\end{equation}
The pump process is characterized by a photon-phonon interaction Hamiltonian of the form
\begin{eqnarray}
\label{Ham}
\mathcal{H}  =  \sum_{j, j'=-J}^{J} \big[ g^1_{j, j'} \,  \mu_{\text{d}}  \big(\hat{a}_{xj}^{\dagger} \, \hat{a}_{yj'} \, \hat{b}^{\dagger} + \hat{a}_{xj} \, \hat{a}_{yj'}^{\dagger} \, \hat{b} \big) +  \, g^2_{j, j'} \,  \mu_{\text{s}}   \big( \hat{a}_{xj}^{\dagger} \, \hat{a}_{yj'} \, (\hat{b}^{\dagger})^2 + \hat{a}_{xj} \, \hat{a}_{yj'}^{\dagger} \, \hat{b}^{2} \big)\big] \ ,
\end{eqnarray}
where $\mu_{\text{d}}$ and $\mu_{\text{s}}$ are coupling constants, $2J +1$ is the total number of modes within a mode-locked optical pulse, and the functions $g^{\ell}_{j, j'}$ take into account the relations between the frequencies of the involved fields,
\begin{equation}
\nonumber
g^{\ell}_{j, j'} =
\left\{
\begin{array}{rl}
1 & \mbox{if } j' = j + \frac{\ell \Omega}{\delta} \\
0 & \mbox{elsewhere},
\end{array}
\right.\quad \ell=1, 2\ .
\\
\end{equation}
\\
As stated in the main text, it should be noted that while the linear term involves only the creation of a phonon in a single mode at null momentum $\mathbf{k}$, the quartic term are not limited to $\mathbf{k}=0$ and one should integrate over the entire optical phonon dispersion including processes where the momentum conservation is guaranteed by the creation of optical phonons with opposite momenta \cite{dorner80, scott68}. In our effective Hamiltonian  we include only a single phonon mode at $\mathbf{k}=0$. This assumption is made in view of the fact that, in the performed experiments, the probing process is limited to the linear regime so that phonons at $\mathbf{k}\neq 0$ will not affect the observed photon number fluctuations in this configuration.

Initially, the sample is in thermal equilibrium and it is described by a thermal phonon state $\hat{\rho}_{\beta}$, at inverse
temperature $\beta$. The Hamiltonian in \eqref{Ham} generates an impulsive change of the initial photon-phonon state $\ket{\bar{\nu}}\bra{\bar{\nu}}\, \otimes \hat{\rho}_{\beta}$ given by
\begin{equation}
\label{pump2}
\hat{\rho}^{\bar{\nu}}=\mathcal{U}\, (\ket{\bar{\nu}}\bra{\bar{\nu}}\, \otimes \hat{\rho}_{\beta}) \, \mathcal{U}^{\dagger}=\ket{\bar{\nu}}\bra{\bar{\nu}}\,\otimes\,\mathcal{U}_{\bar{\nu}}\,\hat{\rho}_\beta\,\mathcal{U}^\dag_{\bar{\nu}}\ ,
\end{equation}
where, because of the high intensity of the pump laser beam, we have adopted the mean field approximation and substituted the photon annihilation and creation operators by the scalar amplitudes $\nu$ and $\nu^*$ and replaced  $\mathcal{U}$ with
\begin{eqnarray}
\label{pump3a}
&&\hskip-1cm
\mathcal{U}_{\bar{\nu}}=\text{exp}\{- i  [c_1\,\hat b^\dag\,+\,c_1^*\,\hat b\,+\,c_2\,(\hat b^\dag)^2\,+\,c_2^*\,\hat b^2]\}\\
&&\hskip-1cm
\label{pump3b}
c_1=\mu_{\text{d}}\,\sum_{j, j'=-J}^{J} g^1_{j, j'} \,{\nu}_{xj}^{*} \, {\nu}_{yj'}\\
&&\hskip-1cm
\label{pump3c}
c_2=\mu_{\text{s}} \sum_{j, j'=-J}^{J}\,g^2_{j, j'} \, {\nu}_{xj}^{*} \, {\nu}_{yj'} \ .\end{eqnarray}
The pump thus prepares the relevant phonon degree of freedom in a state $\hat{\rho}_{\text{II}}^{\bar \nu}$ which is obtained from $\hat{\rho}^{\bar \nu}$ by tracing over the photon degrees of freedom:
\begin{equation}
\label{pump4}
\hat{\rho}_{\text{II}}^{\bar \nu}=\text{Tr}_{\text{I}} ( \,\hat{\rho}^{\bar \nu})=\mathcal{U}_{\bar{\nu}}\,\hat{\rho}_\beta\,\mathcal{U}^\dag_{\bar{\nu}}\ ,
\end{equation}
where $\text{II}$ and $\text{I}$ refer to the phonon and photon system, respectively.

The linear contribution in the phonon operators is responsible for the displacement of $\hat b$ and $\hat b^\dag$, while the quadratic one accounts for their multiplication by hyperbolic functions and thus for the possible squeezing of the corresponding quadratures \cite{scully}:
\begin{eqnarray}
\label{squeez1}
\mathcal{U}_{\bar \nu}^\dag\,\begin{pmatrix}\hat b\cr\hat b^\dag\end{pmatrix}\,\mathcal{U}_{\bar \nu}= \mathbf{S} \,\begin{pmatrix}\hat b\cr\hat b^\dag\end{pmatrix}\,+\,\frac{1}{2|c_2|^2}(\mathbf{S}-1)\begin{pmatrix}c_1^* c_2\cr c_1 c_2^*\end{pmatrix}\\
\mathbf{S}=\begin{pmatrix}\cosh(2|c_2|)&-{\rm e}^{i(\phi+\frac{\pi}{2})} \sinh(2|c_2|) \cr
-{\rm e}^{-i(\phi+\frac{\pi}{2})} \sinh(2|c_2|)&\cosh(2|c_2|)\end{pmatrix}\ ,
\label{squeez2}
\end{eqnarray}
where $c_2=|c_2|{\rm e}^{i\phi}$. In order to write the squeezing matrix $\mathbf{S}$ in the standard formalism \cite{scully}, we can define for convenience a complex squeezing parameter
\begin{equation}
\label{squeezingPar}
\xi = r {\rm e}^{i \psi}, \qquad \text{where} \qquad  r = 2 |c_2| = 2 \,  |\mu_{\text{s}}| \sum_{j, j'=-J}^{J}\,g^2_{j, j'} \, {\nu}_{xj}^{*} \, {\nu}_{yj'}, \qquad \text{and} \qquad \psi = \phi + \frac{\pi}{2}.
\end{equation}
Notice that the squeezing parameter amplitude $r$ depends linearly on the intensity of the pump pulse and on the squeezing coupling constant $\mu_{\text{s}}$ which weights the non linear term in the interaction Hamiltonian and models the material properties involved in the process.

The variance of the quadrature operator $\hat B=\frac{\hat b+\hat b^\dag}{\sqrt{2}}$ with respect to the state $\hat{\rho}_{\text{II}}^{\bar \nu}$ is given by
\begin{eqnarray}
\label{squeez3}
\Delta^2_{\hat{\rho}_{\text{II}}^{\bar \nu}}\hat B=\text{Tr}_{\text{II}}\left(\hat{\rho}_{\text{II}}^{\bar\nu}\hat B^2\right)-\left(\text{Tr}_{\text{II}}\left(\hat{\rho}_{\text{II}}^{\bar \nu}\hat B\right)\right)^2 = \frac{1}{2}\coth\left({\frac{\beta\Omega}{2}}\right) \left[ \cosh(2 r ) - \sinh(2r) \cos{\psi}\right]\ .
\label{varqp}
\end{eqnarray}
Then, for $\psi = 0$ and $r$ large enough, one can make $\Delta^2_{\hat{\rho}_{\text{II}}^{\bar \nu}}\hat B$ smaller than $1/2$ which is
the shot noise variance of $\hat B$ with respect to the vacuum state $\vert 0\rangle$ such that $\hat b\vert 0\rangle=0$.
\smallskip

The photoexcited phonon state $\hat{\rho}^{\bar\nu}_{\text{II}}$ then undergoes a dissipative dynamics that effectively takes into account the interaction of the phonons with their environment until, after a delay time $\tau$, the target is
hit by the probe laser beam.
The phonon dynamics is considered to be that of an open quantum system in weak interaction with a large heat bath that will eventually drive the time-evolving phonon density matrix $\hat{\rho}^{\bar{\nu}}_{\text{II}}(t)=\hat{\rho}_{\text{b}}(t)$ to a thermal state $\hat{\rho}_{\beta'}$ at temperature $T'$ larger than that of the pre-pump phonon state: $\beta'\leq\beta$.
Such a relaxation process is described by a master equation \cite{alicki, breuer} for the phonon density matrix $\hat{\rho}_{\text{b}}(t)$ of the form $\partial_t\hat{\rho}_{\text{b}}(t)=\mathbb{L}[\hat{\rho}_{\text{b}}(t)]$, where the generator of the time evolution is given by
\begin{eqnarray}
\nonumber
\mathbb{L}[\hat{\rho}_{\text{b}}(t)]&=&-i\Big[\Omega\,\hat b^\dag \hat b\,,\,\hat{\rho}_{\text{b}}(t)\Big]\\
\nonumber
&+&\lambda\,(1+n')\,\Big(\hat b\,\hat{\rho}_{\text{b}}(t)\,\hat b^\dag-\frac{1}{2}\Big\{\hat b^\dag \hat b\,,\,\hat{\rho}_{\text{b}}(t)\Big\}\Big)\\
&+&
\lambda\,n'\,\Big(\hat b^\dag\,\hat{\rho}_{\text{b}}(t)\,\hat b-\frac{1}{2}\Big\{\hat b \hat b^\dag\,,\,\hat{\rho}_{\text{b}}(t)\Big\}\Big)\ ,
\label{TDG}
\end{eqnarray}
where
$\displaystyle n'=\frac{1}{{\rm e}^{\beta'\Omega}-1}$ ($> n=\frac{1}{{\rm e}^{\beta\Omega}-1}$), while $\lambda$ is a coupling constant sufficiently small so that the non-negligible presence of the environment can nonetheless be accounted for, in the so-called weak-coupling limit regime \cite{alicki}, by a master equation of the above type.

The first term of $\mathbb{L}$ generates the rotation in time of the phonon mode phase at its own eigenfrequency. The second two contributions consist of a so-called noise term $\hat b\,\hat{\rho}_{\text{b}}(t)\,\hat b^\dag$, respectively $\hat b^\dag\,\hat{\rho}_{\text{b}}(t)\,\hat b$ that has the property of transforming pure states into mixed states and of a dissipative term
$-\frac{1}{2}\Big\{\hat b^\dag \hat b\,,\,\hat{\rho}_{\text{b}}(t)\Big\}$, respectively $-\frac{1}{2}\Big\{\hat b \hat b^\dag\,,\,\hat{\rho}_{\text{b}}(t)\Big\}$. These terms counterbalance the noise by keeping the trace of the time-evolving state $\hat{\rho}_{\text{b}}(t)$, and thus the overall probability, constant in time. The anti-commutators can be incorporated into the Hamiltonian as anti-Hermitian contributions responsible for exponential time relaxation.
The structure of $\mathbb{L}$ is such that the generated time-evolution maps, formally $\gamma_t=\exp(t\mathbb{L})$, compose as a forward-in-time semigroup: $\gamma_t\circ\gamma_s=\gamma_s\circ\gamma_t=\gamma_{t+s}$ for all $s,t\geq 0$. Moreover, $\hat{\rho}_{\text{b}}(t)=\gamma_t[\hat{\rho}_{\text{b}}]$ can be explicitly computed for any initial phonon state $\hat{\rho}_{\text{b}}$; all initial states are eventually driven to a unique invariant state satisfying $\mathbb{L}[\hat{\rho}_{\text{b}}]=0$ that is given by the thermal state $\hat{\rho}_{\beta'}$.
\smallskip

Finally, the probe process is again described by the Hamiltonian in equation \eqref{Ham}. However, the corresponding impulsive unitary operator $\mathcal{U}=\exp{(-i \mathcal{H})}$ now acts on a photon-phonon state of the form
$\ket{\bar{\alpha}}\bra{\bar{\alpha}}\,\otimes \, \hat{\rho}^{\bar{\nu}}_{\text{II}} (\tau)$.
Here, $\ket{\bar{\alpha}}\bra{\bar{\alpha}}$ is the multi-mode coherent state associated with the probe laser beam which contains $x$ and $y$ polarized components and is much less intense than the pump one, while $\hat{\rho}_{\text{II}}^{\bar{\nu}}(\tau)$ is the phonon state dissipatively evolved up to the delay time $\tau$ between pump and probe. Differently from the pump process, the lower probe intensity allows one to neglect in $\mathcal{H}$ the quartic terms responsible for the squeezing effects. Moreover, we can apply the mean field approximation only to the field operators with $x$ polarization, since these probe components are much more intense than those polarized along $y$. Then, by replacing $\hat a_{xj}$ and $\hat a_{xj}^\dag$ by $\alpha_{xj}$ and
$\alpha_{xj}^*$ the probe process is described by
\begin{equation}
\label{probe1}
\mathcal{U}_{\bar{\alpha'}} = \text{exp}\{-i \lVert \bar\alpha'\rVert \big(\hat A(\bar\alpha') \, \hat{b}^{\dagger} + \hat A^\dag(\bar\alpha') \, \hat{b} \big)\} \ ,
\end{equation}
where $\hat A(\bar\alpha')$ is the collective photon annihilation operator
\begin{equation}
\label{probe2}
\hat A(\bar\alpha')=\frac{1}{\lVert \bar\alpha'\rVert}\sum_{j=-J}^J(\alpha'_j)^{*}\,\hat a_{yj}\ ,\quad \alpha'_j=\mu_{\text{d}} \,\sum_{j'=-J}^{J}\, g^1_{j', j} \,{\alpha}_{xj}\ .
\end{equation}
Then, the probe process affects an initial state $\ket{\bar{\alpha}_y}\bra{\bar{\alpha}_y}\,\otimes \, \hat{\rho}^{\bar{\nu}}_{\text{II}} (\tau)$, where $\ket{\bar{\alpha}_y} = \ket{\alpha_{y\, -J}} \otimes \cdots \otimes \ket{\alpha_{y\, J}}$ is the coherent state involving only the $y$ polarization components such that $\hat a_{yj} \ket{\bar{\alpha}_y} = \alpha_{yj} \ket{\bar{\alpha}_y}$.\\
Notice that, unlike in \eqref{pump2}, $\mathcal{U}_{\bar{\alpha'}}$ acts on the photon-phonon state as a whole and transforms it into
\begin{equation}
\label{probe3}
\mathcal{U}_{\bar{\alpha'}} \, \ket{\bar{\alpha}_y}\bra{\bar{\alpha}_y} \, \otimes \, \hat{\rho}^{\nu}_{\text{II}}(\tau) \, \mathcal{U}^{\dagger}_{\bar{\alpha'}}\ .
\end{equation}
This allows for  the quantum features of the phonon state and of its dynamics to be transcribed onto the emitted photon state
\begin{equation}
\label{probe1a}
\hat{\rho}_{\text{I}} (\tau) = \text{Tr}_{\text{II}}\, \left(\mathcal{U}_{\bar{\alpha'}} \, \ket{\bar{\alpha}_y}\bra{\bar{\alpha}_y} \, \otimes \, \hat{\rho}^{\nu}_{\text{II}}(\tau) \, \mathcal{U}^{\dagger}_{\bar{\alpha'}}\right)\ .
\end{equation}
Unlike in the semi-classical theoretical approaches to pump and probe experiments attempted so far, one can here confront the experimental data not only with the scattered probe beam intensity, namely with the mean photon number $\braket{\hat N_y}_\tau$, where $\hat N_y=\hat A^\dag(\bar\alpha') \hat A(\bar\alpha')$ and
\begin{equation}
\label{probe4}
\braket{\hat N_y}_\tau=\text{Tr}\left(\hat N_y\,\hat{\rho}_{\text{I}} (\tau)\right)\ ,
\end{equation}
but also with its variance $\Delta^2_\tau\hat N_y=\braket{\hat N^2_y}_\tau-\braket{\hat N_y}_\tau^2$.\\
Then one uses that
\begin{eqnarray}
U^\dag_{\bar\alpha'} \, \hat N_y \, U_{\bar\alpha'} &=& A^\dag(\bar\alpha') \hat A(\bar\alpha') \cos^2(\lVert \bar\alpha'\rVert) + \hat b^\dag \hat b \sin^2(\lVert \bar\alpha'\rVert)
 + \frac{i}{2} \sin(2 \lVert \bar\alpha'\rVert) \big( A(\bar\alpha') \hat b^\dag + A^\dag(\bar\alpha') \hat b  \big),
\label{Ncoll}
\end{eqnarray}
where, given the experimental conditions effectively described by the model,
it is plausible to set all
amplitudes $\alpha_{xj}=\alpha_x $ and $\alpha_{yj}=\alpha_y = |\alpha_y|\text{exp}(i \theta_y)$, in which case
\begin{eqnarray}
\label{coeff}
\alpha'_j = \mu_{\text{d}} \alpha_x = |\mu_{\text{d}} \alpha_x| e^{i \theta'}, \qquad \text{and} \qquad \lVert \bar\alpha'\rVert = \sqrt{K} |\mu_{\text{d}} \alpha_x| \, ,
\end{eqnarray}
where $K=2J+1$ is the total number of modes within a mode-locked optical pulse.\\
By denoting with  $I_y = K |\alpha_y|^2$  the pulse intensity for the $y$ polarization and using that
\begin{equation}
\hat A(\bar\alpha') \ket{\bar\alpha_y} = \sqrt{I_y} e^{-i(\theta'-\theta_y)} \ket{\bar\alpha_y},
\end{equation}
one explicitly computes:
\begin{eqnarray}
\braket{\hat N_y}_\tau &=& I_y \cos^2(\lVert \bar\alpha'\rVert) + \sin^2(\lVert \bar\alpha'\rVert) \braket{\hat b^\dag \hat b}_\tau \, + \, \frac{i}{2} \sqrt{I_y} \, \sin(2 \lVert \bar\alpha'\rVert) \, \big( e^{-i (\theta' - \theta_y)} \braket{\hat b^\dag}_\tau - e^{i (\theta' - \theta_y)} \braket{\hat b}_\tau\big)\,,
\label{probe5}
\end{eqnarray}
where
$\braket{\hat O}_\tau=\text{Tr}_{\text{II}}\left(\hat{\rho}^{\bar \nu}_{\text{II}}(\tau)\,\hat O\right)$ is the expectation value of any phonon operator $\hat O$ with respect to the phonon state $\hat{\rho}_{\text{II}}^{\bar \nu}(\tau)$.\\
Despite its complicated expression, we report also the number variance $\Delta^2_\tau\hat N_y$ predicted by the model, as $\braket{\hat N_y}_\tau$ and $\Delta^2_\tau\hat N_y$ are the quantities computed numerically in the main text and compared with the experimental data:
\begin{eqnarray}
\nonumber
\Delta^2_\tau\hat N_y &=& I_y \cos^4( \lVert \bar\alpha'\rVert)  +\sin^4(\lVert \bar\alpha'\rVert)\left(\braket{(\hat b^\dag\hat b)^2}_\tau - \braket{\hat b^\dag\hat b}_\tau^2\right) + \sin^2(\lVert \bar\alpha'\rVert)\cos^2(\lVert \bar\alpha'\rVert)\,  \braket{\hat b^\dag\hat b}_\tau \\
\nonumber
        &-& I_y \sin^2(\lVert \bar\alpha'\rVert)\cos^2(\lVert \bar\alpha'\rVert)\,  \left[e^{-2i (\theta' - \theta_y)} \left( \braket{(\hat b^\dag)^2}_\tau - \braket{\hat b^\dag}_\tau^2 \right) + e^{2i(\theta' - \theta_y)} \left(\braket{\hat b^2}_\tau - \braket{\hat b}_\tau^2\right)\right]\\
\nonumber
        &+& I_y \sin^2(\lVert \bar\alpha'\rVert)\cos^2(\lVert \bar\alpha'\rVert)\, \left( 2\braket{\hat b^\dag\hat b}_\tau + 1 -  2 \braket{\hat b^\dag}_\tau \braket{\hat b}_\tau \right)\\
\nonumber
        &+& i \sqrt{I_y} \, \sin(\lVert \bar\alpha'\rVert)\cos^3(\lVert \bar\alpha'\rVert)\, \left(e^{-i (\theta' - \theta_y)} \braket{\hat b^\dag}_\tau - e^{i (\theta' - \theta_y)} \braket{\hat b}_\tau   \right)\\
\nonumber
        &+& i \sqrt{I_y} \, \sin^3(\lVert \bar\alpha'\rVert)\cos(\lVert \bar\alpha'\rVert)\, \left[  2e^{-i (\theta' - \theta_y)} \left( \braket{(\hat b^\dag)^2\hat b}_\tau - \braket{\hat b^\dag\hat b}_\tau \braket{\hat b^\dag}_\tau + \frac{1}{2} \braket{\hat b^\dag}_\tau \right) \right. \\
\nonumber
         &-& \left.2e^{i (\theta' - \theta_y)} \left( \braket{\hat b^\dag\hat b^2}_\tau - \braket{\hat b^\dag\hat b}_\tau\braket{\hat b}_\tau  + \frac{1}{2}\braket{\hat b}_\tau  \right)  \right] .\\
\label{var}
\end{eqnarray}

The phononic correlation functions involving $\hat{b}$ and $\hat{b}^\dag$ contribute with oscillations at frequency $\Omega$ while those involving $\hat{b}^2$ and $\hat{b}^{\dag 2}$ give rise to $2 \Omega$ oscillations. Collecting the corresponding coefficients one finds the following amplitude for the $2 \Omega$ oscillating components:
\begin{equation}
\label{A2omega}
\left|A_{2 \Omega} (\tau)\right| = \frac{I_y  (1 + 2 n') }{8} {\rm e}^{-\lambda \tau}  \sin^2(2 \lVert \bar\alpha'\rVert)\sinh(2  r)\ ,
\end{equation}
where the amplitude of the squeezing parameter $r = 2 |c_2| = 2 K |\mu_{\text{s}}| |\nu|^2$ is obtained from \eqref{squeezingPar} and \eqref{pump3c} by putting all pump amplitudes equal to $\nu$. Moreover we take $\lambda$ to comply with the observed oscillation time-scale and the time $\tau>0$ such that $\lambda \tau \ll 1$.

In the last figure in the main text we have shown a fit of the experimental results for different pump intensities with the functional behaviour of $A_{2 \Omega}$ predicted by the model in equation \eqref{A2omega}. We found an optimal value of the coupling parameter $\mu_{\text{s}}$ for which the model agrees with the experiments. We used such a value for computing the amplitude $r$ of the squeezing parameter (defined in equation \eqref{pump3c}) for all the experimental pump fluences. In particular  $|\nu|^2$  is the number of photons per unit cell per pulse. We then computed the uncertainties in the position and momentum phonon operators as in equation \eqref{varqp}. The results reported in the main text unveil photo-excited thermal squeezed vibrational states.

For completeness we also report the explicit time evolution of $\Delta^2_\tau\hat N_y$ in terms of both the amplitude $A_{2\Omega}$ of the $2 \Omega$ frequency component and the amplitude $A_\Omega$ of the fundamental frequency component:
\begin{equation}
\Delta^2_\tau\hat N_y = A_{0} (\tau)+ A_{\Omega} (\tau) e^{i\Omega \tau}+ A^{\ast}_{\Omega} (\tau) e^{-i\Omega \tau}+ A_{2\Omega} (\tau) e^{2i\Omega \tau}+ A^{\ast}_{2\Omega} (\tau) e^{-2i\Omega \tau},
\end{equation}
where the explicit expression for $\left|A_\Omega (\tau)\right|$ is:
\begin{eqnarray}
\label{Aomega}
\left|A_{\Omega} (\tau) \right| &=& \frac{1}{2}\sqrt{I_{y}} \, |z| \,  e^{-\lambda \tau/2} \left|  2 e^{-\lambda \tau} \sin(2\lVert \bar\alpha'\rVert) \sin^{2}(\lVert \bar\alpha'\rVert) \left[ \left(1+n'+n - (1+2n')\cosh(2r) \right) e^{i \theta_{r}}\cosh(r) \right.  \right.\\
\nonumber
 &+& \left. \left( n'-n + (1+2n')\cosh(2r) \right)  e^{2i\theta_{z}}\sinh(r) \right] \\
\nonumber
 &-& \left. \sin(2\lVert \bar\alpha'\rVert)\left( 1+2n \sin^{2}(\lVert \bar\alpha'\rVert) \right)\left(  e^{i\theta_{r}}\cosh(r)-  e^{2i \theta_{z}}\sinh(r) \right) \right|\, ,
\end{eqnarray}
where $|z|$ is the corresponding photo-exited displacement in the phonon and $\theta_{z}$ its phase.\\
We stress that, if the pump pulse does not generate squeezed phonons, vanishing squeezing parameter ($r\rightarrow 0$), the amplitudes of the two frequency components become,
\begin{eqnarray}
\left|A_{\Omega} (\tau) \right| &=& \dfrac{1}{2}\sqrt{I_{y}} \, z  \, e^{-\lambda \tau/2} \left| \sin(2\lVert \bar\alpha'\rVert) + 2\sin(2\lVert \bar\alpha'\rVert)\sin^{2}
(\lVert \bar\alpha'\rVert) \left( n - (n - n') e^{-\lambda \tau} \right)\right| \, , \\
\nonumber
&&\\
\nonumber
\left| A_{2\Omega} (\tau)  \right| &=& 0 ,
\end{eqnarray}
indicating the absence of the $2 \Omega$ frequency component in the variance in absence of phonon squeezing.

From equations \eqref{Aomega} and \eqref{A2omega} one can notice that the damping constant $\lambda$, characterizing the dissipative phonon time evolution between the excitation and the probing process,  contributes differently to  $A_{2 \Omega}$ and to  $A_{\Omega}$, giving rise to different decay times for the two components and reproducing the experimental results.

\renewcommand{\refname}{\textbf{\large Supplemenary References}}